\documentclass[25pt]{article}
\usepackage{amsfonts,amsmath,amssymb,amsthm,epsfig,mathrsfs,subfigure}
\usepackage{setspace}
\theoremstyle{definition}

\theoremstyle{plain}

\title{Riemann-Hilbert method for $N$-soliton solution of the coupled Hirota system in nonlinear fiber\footnote{This work was supported by the National Natural Science Foundation of China (Grant Nos. 61072147 and 11271008).}}
\author{Zhou-Zheng Kang$^{1,2}$ and Tie-Cheng Xia$^{1}$\footnote{Corresponding author. E-mail: xiatc@shu.edu.cn.} \\
1. Department of Mathematics, Shanghai University, Shanghai 200444, China;\\
2. College of Mathematics, Inner Mongolia University for Nationalities,\\ Tongliao 028043, China\\}
\date{}
\begin{document}
\sloppy \maketitle
\begin{abstract}
ժҪUnder investigation in this work is the coupled Hirota system arising in nonlinear fiber. The spectral analysis of the Lax pair is first carried out and a Riemann-Hilbert problem is described. Then in the framework of the obtained Riemann-Hilbert problem with the reflectionless case, $N$-soliton solution is presented for the coupled Hirota system. Finally, via proper choices of the involved parameters, a few plots are made to exhibit the localized structures and dynamic characteristics of one-soliton solution. \vskip
2mm\noindent\textbf{AMS Subject classification}: 35Q15, 35C08\vskip
2mm\noindent\textbf{Keywords}: Coupled Hirota system; Riemann-Hilbert method; soliton solutions
\end{abstract}
\section{Introduction}
As is well known, investigating soliton solutions to nonlinear evolution equaitons (NLEEs) is of an especially important significance in the study of various nonlinear phenomena arising from fluid dynamics, plasma physics, oceanography, optics, condensed matter physics and so on. Until now,
there have been a number of powerful methods for seeking soliton solutions of NLEEs, some of which include the inverse scattering transformation method [1], the Darboux transformation method [2,3], the B\"{a}cklund transformation method [4], the Riemann-Hilbert approach [5] and the diverse function expansion methods [6,7]. In recent years, the Riemann-Hilbert approach providing a means to explore abundant multi-soliton solutions has been adopted to study many NLEEs [8-15].

In this paper, we are concerned with the coupled Hirota system [16]
\begin{subequations}
\begin{align}
&{{q}_{1t}}+2{{A}_{2}}{{q}_{1xx}}+4k_{1}^{2}{{A}_{2}}\big({{\left| {{q}_{1}} \right|}^{2}}+{{\left| {{q}_{2}} \right|}^{2}}\big){{q}_{1}}\nonumber\\
&\quad\ -\varepsilon \big[{{q}_{1xxx}}+3k_{1}^{2}\big({{\left| {{q}_{1}} \right|}^{2}}+{{\left| {{q}_{2}} \right|}^{2}}\big){{q}_{1x}}+3k_{1}^{2}{{q}_{1}}(q_{1}^{*}{{q}_{1x}}+q_{2}^{*}{{q}_{2x}}) \big]=0,\label{1a} \\
&{{q}_{2t}}+2{{A}_{2}}{{q}_{2xx}}+4k_{1}^{2}{{A}_{2}}\big({{\left| {{q}_{1}} \right|}^{2}}+{{\left| {{q}_{2}} \right|}^{2}}\big){{q}_{2}}\nonumber\\
&\quad\ -\varepsilon \big[{{q}_{2xxx}}+3k_{1}^{2}\big({{\left| {{q}_{1}} \right|}^{2}}+{{\left| {{q}_{2}} \right|}^{2}}\big){{q}_{2x}}+3k_{1}^{2}{{q}_{2}}(q_{1}^{*}{{q}_{1x}}+q_{2}^{*}{{q}_{2x}})\big]=0,\label{1b}
\end{align}
\end{subequations}
which describes the pulse propagation in a coupled fiber with higher-order
dispersion and self-steepening. Here $q_{1}$ and $q_{2}$ represent the complex amplitude of the pulse envelope,
the subscripts of $q_{1}$ and $q_{2}$ denote the partial derivatives with respect to the
scaled spatial coordinate $x$ and time coordinate $t$ correspondingly,
while $\varepsilon$, $k_{1}$, $A_{2}$ are three real parameters, and $\ast$ means complex conjugate.
To the best of our knowledge, the construction of $N$-soliton solution for the coupled Hirota system (1) has not been reported based on the Riemann-Hilbert approach, which is the motivation of our paper.

The rest of the paper is organized as follows. Section 2 is devoted to analyzing the spectral problem of the Lax pair and putting forward a Riemann-Hilbert problem for the system (1).
In Section 3, through solving the resulting Riemann-Hilbert problem corresponding to the reflectionless case, $N$-soliton solution to system (1) is generated. Lastly, a short conclusion is given.
\section{Riemann-Hilbert problem}
In this section, our aim is to put forward a Riemann-Hilbert problem for the coupled Hirota system (1),
which will lay the ground for us to derive
$N$-soliton solution in next section. System (1) admits the $3\times3$ Lax pair$^{[16]}$
\begin{subequations}\begin{align}
 & {{\Psi}_{x}}=U\Psi,\tag{2a} \\
 & {{\Psi}_{t}}=V\Psi,\tag{2b}
\end{align}
\end{subequations}
where $\Psi={({{\Psi}_{1}}(x,t),{{\Psi}_{2}}(x,t),{{\Psi}_{3}}(x,t))^\textrm{T}}$ is the spectral function, the symbol $\textrm{T}$ means transpose of the vector, and
\[\begin{aligned}
  & U=\left( \begin{matrix}
   -\frac{i}{2}\zeta  & -{{k}_{1}}{{q}_{1}} & -{{k}_{1}}{{q}_{2}}  \\
   {{k}_{1}}q_{1}^{*} & \frac{i}{2}\zeta  & 0  \\
   {{k}_{1}}q_{2}^{*} & 0 & \frac{i}{2}\zeta   \\
\end{matrix} \right), \\
 & V={{\zeta}^{3}}\left( \begin{matrix}
   \frac{i}{2}\varepsilon  & 0 & 0  \\
   0 & -\frac{i}{2}\varepsilon  & 0  \\
   0 & 0 & -\frac{i}{2}\varepsilon   \\
\end{matrix} \right)+{{\zeta}^{2}}\left( \begin{matrix}
   {{A}_{2}} & \varepsilon {{k}_{1}}{{q}_{1}} & \varepsilon {{k}_{1}}{{q}_{2}}  \\
   -\varepsilon {{k}_{1}}q_{1}^{*} & -{{A}_{2}} & 0  \\
   -\varepsilon {{k}_{1}}q_{2}^{*} & 0 & -{{A}_{2}}  \\
\end{matrix} \right) \\
 &\quad\quad+\zeta \left( \begin{matrix}
   -i\varepsilon k_{1}^{2}\big({{\left| {{q}_{1}} \right|}^{2}}+{{\left| {{q}_{2}} \right|}^{2}}\big) & i\varepsilon {{k}_{1}}{{q}_{1x}}-2i{{A}_{2}}{{k}_{1}}{{q}_{1}} & i\varepsilon {{k}_{1}}{{q}_{2x}}-2i{{A}_{2}}{{k}_{1}}{{q}_{2}}  \\
   i\varepsilon {{k}_{1}}q_{1x}^{*}+2i{{A}_{2}}{{k}_{1}}q_{1}^{*} & i\varepsilon k_{1}^{2}{{\left| {{q}_{1}} \right|}^{2}} & i\varepsilon k_{1}^{2}q_{1}^{*}{{q}_{2}}  \\
   i\varepsilon {{k}_{1}}q_{2x}^{*}+2i{{A}_{2}}{{k}_{1}}q_{2}^{*} & i\varepsilon k_{1}^{2}q_{2}^{*}{{q}_{1}} & i\varepsilon k_{1}^{2}{{\left| {{q}_{2}} \right|}^{2}}  \\
\end{matrix} \right)+\left( \begin{matrix}
   {{B}_{11}} & {{B}_{12}} & {{B}_{13}}  \\
   {{B}_{21}} & {{B}_{22}} & {{B}_{23}}  \\
   {{B}_{31}} & {{B}_{32}} & {{B}_{33}}  \\
\end{matrix} \right).
\end{aligned}\]
Here $\zeta\in \mathbb{C}$ is a spectral parameter. Besides,
\[\begin{aligned}
 & {{B}_{11}}=-2{{A}_{2}}k_{1}^{2}\big({{\left| {{q}_{1}} \right|}^{2}}+{{\left| {{q}_{2}} \right|}^{2}}\big)-\varepsilon k_{1}^{2}({{q}_{1}}q_{1x}^{*}-q_{1}^{*}{{q}_{1x}}+{{q}_{2}}q_{2x}^{*}-q_{2}^{*}{{q}_{2x}}), \\
 & {{B}_{12}}=-\varepsilon {{k}_{1}}{{q}_{1xx}}+2{{A}_{2}}{{k}_{1}}{{q}_{1x}}-2\varepsilon k_{1}^{3}{{q}_{1}}\big({{\left| {{q}_{1}} \right|}^{2}}+{{\left| {{q}_{2}} \right|}^{2}}\big),\\&{{B}_{13}}=-\varepsilon {{k}_{1}}{{q}_{2xx}}+2{{A}_{2}}{{k}_{1}}{{q}_{2x}}-2\varepsilon k_{1}^{3}{{q}_{2}}\big({{\left| {{q}_{1}} \right|}^{2}}+{{\left| {{q}_{2}} \right|}^{2}}\big), \\
 & {{B}_{21}}=\varepsilon {{k}_{1}}q_{1xx}^{*}+2{{A}_{2}}{{k}_{1}}q_{1x}^{*}+2\varepsilon k_{1}^{3}q_{1}^{*}\big({{\left| {{q}_{1}} \right|}^{2}}+{{\left| {{q}_{2}} \right|}^{2}}\big),\\&{{B}_{22}}=-\varepsilon k_{1}^{2}(q_{1}^{*}{{q}_{1x}}-q_{1x}^{*}{{q}_{1}})+2{{A}_{2}}k_{1}^{2}{{\left| {{q}_{1}} \right|}^{2}}, \\
 & {{B}_{23}}=-\varepsilon k_{1}^{2}(q_{1}^{*}{{q}_{2x}}-{{q}_{1x}}{{q}_{2}})+2{{A}_{2}}k_{1}^{2}q_{1}^{*}{{q}_{2}},\\&{{B}_{31}}=\varepsilon {{k}_{1}}q_{2xx}^{*}+2{{A}_{2}}{{k}_{1}}q_{2x}^{*}+2\varepsilon k_{1}^{3}q_{2}^{*}\big({{\left| {{q}_{1}} \right|}^{2}}+{{\left| {{q}_{2}} \right|}^{2}}\big), \\
 & {{B}_{32}}=-\varepsilon k_{1}^{2}(q_{2}^{*}{{q}_{1x}}-{{q}_{1}}q_{2x}^{*})+2{{A}_{2}}k_{1}^{2}q_{2}^{*}{{q}_{1}},\\&{{B}_{33}}=-\varepsilon k_{1}^{2}(q_{2}^{*}{{q}_{2x}}-q_{2x}^{*}{{q}_{2}})+2{{A}_{2}}k_{1}^{2}{{\left| {{q}_{2}} \right|}^{2}}.
\end{aligned}\]

For ease of analysis, we convert the Lax pair (2) into the equivalent form
\begin{subequations}
\begin{align}
 & {{\Psi}_{x}}=\left( \frac{i}{2}\zeta\sigma -{{k}_{1}}Q \right)\Psi,\label{3a} \tag{3a} \\
 & {{\Psi}_{t}}=\left( -\frac{i}{2}{{\zeta}^{3}}\varepsilon \sigma -{{\zeta}^{2}}{{A}_{2}}\sigma +\tilde{Q} \right)\Psi, \tag{3b}
\end{align}
\end{subequations}
where $\sigma=\text{diag}(-1,1,1),\tilde{Q}={{\zeta}^{2}}\varepsilon {{k}_{1}}Q+\zeta Q_{0}+B,$ and
\[\begin{aligned}
 & Q=\left( \begin{matrix}
   0 & {{q}_{1}} & {{q}_{2}}  \\
   -q_{1}^{*} & 0 & 0  \\
   -q_{2}^{*} & 0 & 0  \\
\end{matrix} \right),\quad B=\left( \begin{matrix}
   {{B}_{11}} & {{B}_{12}} & {{B}_{13}}  \\
   {{B}_{21}} & {{B}_{22}} & {{B}_{23}}  \\
   {{B}_{31}} & {{B}_{32}} & {{B}_{33}}  \\
\end{matrix} \right),\\&Q_{0}=\left( \begin{matrix}
   -i\varepsilon k_{1}^{2}\big({{\left| {{q}_{1}} \right|}^{2}}+{{\left| {{q}_{2}} \right|}^{2}}\big) & i\varepsilon {{k}_{1}}{{q}_{1x}}-2i{{A}_{2}}{{k}_{1}}{{q}_{1}} & i\varepsilon {{k}_{1}}{{q}_{2x}}-2i{{A}_{2}}{{k}_{1}}{{q}_{2}}  \\
   i\varepsilon {{k}_{1}}q_{1x}^{*}+2i{{A}_{2}}{{k}_{1}}q_{1}^{*} & i\varepsilon k_{1}^{2}{{\left| {{q}_{1}} \right|}^{2}} & i\varepsilon k_{1}^{2}q_{1}^{*}{{q}_{2}}  \\
   i\varepsilon {{k}_{1}}q_{2x}^{*}+2i{{A}_{2}}{{k}_{1}}q_{2}^{*} & i\varepsilon k_{1}^{2}q_{2}^{*}{{q}_{1}} & i\varepsilon k_{1}^{2}{{\left| {{q}_{2}} \right|}^{2}}
\end{matrix} \right).\end{aligned}\]

Suppose that the potential functions $q_{1}$ and $q_{2}$ in the Lax pair (3) decay to zero sufficiently fast as $x\rightarrow\pm\infty$. From (3), we see that $\Psi \propto{{\text{e}}^{\frac{i}{2}\zeta \sigma x+\left( -\frac{i}{2}{{\zeta}^{3}}\varepsilon -{{\zeta}^{2}}{{A}_{2}} \right)\sigma t}}$ when $x\rightarrow\pm\infty$,
which motivates us to introduce the following transformation
\begin{equation*}
\Psi=\mu{{\text{e}}^{\frac{i}{2}\zeta \sigma x+\left( -\frac{i}{2}{{\zeta}^{3}}\varepsilon -{{\zeta}^{2}}{{A}_{2}} \right)\sigma t}}.
\end{equation*}
Using this transformation, the Lax pair (3) is then turned into the desired form
\begin{subequations}
\begin{align}
 & {{\mu}_{x}}=\frac{i}{2}\zeta [\sigma ,\mu]+{{U}_{1}}\mu, \tag{4a}\\
 & {{\mu}_{t}}=\left( -\frac{i}{2}{{\zeta}^{3}}\varepsilon -{{\zeta}^{2}}{{A}_{2}} \right)[\sigma ,\mu]+\tilde{Q}\mu. \tag{4b}
\end{align}
\end{subequations}
Here $[\sigma ,\mu]\equiv\sigma \mu-\mu\sigma$ denotes the commutator and $U_{1}=-k_{1}Q$.

Now we investigate the direct scattering.
Since the analysis will take place at a fixed time, the $t$-dependence will be suppressed.
As for the spectral problem (4a), we introduce its two matrix Jost solutions $\mu_{\pm}$ written as a collection of columns
$$
{\mu_{-}}=({{[{\mu_{-}}]}_{1}},{{[{\mu_{-}}]}_{2}},{{[{\mu_{-}}]}_{3}}),\quad {\mu_{+}}=({{[{\mu_{+}}]}_{1}},{{[{\mu_{+}}]}_{2}},{{[{\mu_{+}}]}_{3}})\eqno(5)
$$
with the asymptotic conditions
\begin{subequations}
\begin{align}
&{\mu_{-}}\to \mathbb{I},\quad x\to -\infty , \tag{6a}\\
&{\mu_{+}}\to \mathbb{I},\quad x\to +\infty . \tag{6b}
\end{align}
\end{subequations}
Here the subscripts of $\mu$ mean which end of the $x$-axis the boundary conditions are set, and $\mathbb{I}$ is a identity matrix of size 3. As a matter of fact, the solutions ${\mu_{\pm }}$ are uniquely determined by the integral equations of Volterra type
\begin{subequations}
\begin{align}
&{\mu_{-}}=\mathbb{I}+\int_{-\infty }^{x}{{{\text{e}}^{\frac{i}{2}\zeta\sigma (x-\xi)}}{{U}_{1}}(\xi){{\mu}_{-}}(\xi,\zeta){{\text{e}}^{-\frac{i}{2}\zeta \sigma (x-\xi)}}\text{d}\xi},\tag{7a}\\
&{\mu_{+}}=\mathbb{I}-\int_{x}^{+\infty }{{{\text{e}}^{\frac{i}{2}\zeta\sigma (x-\xi)}}{{U}_{1}}(\xi){{\mu}_{+}}(\xi,\zeta){{\text{e}}^{-\frac{i}{2}\zeta \sigma (x-\xi)}}\text{d}\xi}.\tag{7b}
\end{align}
\end{subequations}
After some calculations we see from (7) that ${{[{\mu_{+}}]}_{1}},{{[{\mu_{-}}]}_{2}},{{[{\mu_{-}}]}_{3}}$ are analytic for $\zeta \in {\mathbb{C}^{-}}$ and continuous for $\zeta \in {\mathbb{C}^{-}}\cup \mathbb{R}$, whereas ${{[{\mu_{-}}]}_{1}},{{[{\mu_{+}}]}_{2}},{{[{\mu_{+}}]}_{3}}$ are analytic for $\zeta\in {\mathbb{C}^{+}}$ and continuous for $\zeta \in {\mathbb{C}^{+}}\cup \mathbb{R}$, where ${\mathbb{C}^{-}}$ and ${\mathbb{C}^{+}}$ are respectively the lower and upper half $\zeta$-plane.

Next we focus on studying the properties of $\mu_{\pm}$.
Actually, it is indicated owing to $\text{tr}(Q)=0$ that the determinants of ${\mu_{\pm }}$ are independent of the variable $x$. Evaluating $\det {\mu_{-}}$ at $x=-\infty$ and $\det {\mu_{+}}$ at $x=+\infty$, we find $\det {\mu_{\pm }}=1$ for $\zeta \in \mathbb{R}.$
Additionally, ${\mu_{-}}E$ and ${\mu_{+}}E$ are both the matrix solutions of the original spectral problem (3a),
where $E={{\text{e}}^{\frac{i}{2}\zeta \sigma x}}$,
they are linearly dependent
$$
{\mu_{-}}E={\mu_{+}}ES(\zeta),\quad \zeta \in \mathbb{R}.\eqno(8)
$$
Here $S(\zeta)={{({{s}_{kj}})}_{3\times 3}}$ is a scattering matrix, and
$\det S(\zeta)=1.$
Furthermore, it is found from the property of $\mu_{-}$ that ${{s}_{11}}$ admits analytic extension to ${\mathbb{C}^{+}}$ and ${{s}_{kj}}$\ $(k,j=2,3)$ analytically extend to ${\mathbb{C}^{-}}$.

A Riemann-Hilbert problem we are seeking is associated with two matrix functions: one is analytic in ${\mathbb{C}^{+}}$ and the other is analytic in ${\mathbb{C}^{-}}$. Let the first analytic function of $\zeta$ in ${\mathbb{C}^{+}}$ be of the form
$$
{{P}_{1}}(x,\zeta)=({{[{\mu_{-}}]}_{1}},{{[{\mu_{+}}]}_{2}},{{[{\mu_{+}}]}_{3}})(x,\zeta).\eqno(9)
$$
And then, one can expand ${{P}_{1}}$ into the asymptotic series at large-$\zeta$
$$
{{P}_{1}}=P_{1}^{(0)}+\frac{P_{1}^{(1)}}{\zeta}+\frac{P_{1}^{(2)}}{{{\zeta}^{2}}}+O\bigg( \frac{1}{{{\zeta}^{3}}} \bigg),\quad \zeta \to \infty. \eqno(10)
$$
Through carrying expansion (10) into the spectral problem (4a) and equating terms with same powers of $\zeta$, we have
\begin{equation*}
\frac{i}{2}\big[\sigma ,P_{1}^{(1)}\big]+{{U}_{1}}P_{1}^{(0)}=P_{1x}^{(0)},\quad
\frac{i}{2}\big[\sigma ,P_{1}^{(0)}\big]=0,
\end{equation*}
which yields $P_{1}^{(0)}=\mathbb{I}$, i.e., ${{P}_{1}}\to \mathbb{I}$ as $ \zeta \in {\mathbb{C}^{+}}\to \infty .$

To set up a Riemann-Hilbert problem, there is still the analytic counterpart of $P_{1}$ in ${\mathbb{C}^{-}}$ to be constructed.
Noting that the adjoint scattering equation of (4a) is of the form
$$
{{K}_{x}}=\frac{i}{2}\zeta[\sigma ,K]-K{{U}_{1}}.\eqno(11)
$$
The inverse matrices of ${\mu_{\pm }}$ can be partitioned into rows
$$
\mu_{\pm}^{-1}=\left( \begin{matrix}
   {[\mu_{\pm}^{-1}]^{1}}  \\
   {[\mu_{\pm}^{-1}]^{2}}  \\
   {[\mu_{\pm}^{-1}]^{3}}  \\
\end{matrix} \right)\eqno(12)
$$
obeying the boundary conditions $\mu_{\pm}^{-1}\rightarrow\mathbb{I}$ as $x\rightarrow\pm\infty$.
And it is easily known that $\mu_{\pm}^{-1}$ solve the adjoint equation (11).
From (8) we know
$$
{{E}^{-1}}\mu_{-}^{-1}=R(\zeta){{E}^{-1}}\mu_{+}^{-1}\eqno(13)
$$
with $R(\zeta)={{({{r}_{kj}})}_{3\times 3}}$ as the inverse matrix of $S(\zeta)$.
Consequently, the matrix function ${{P}_{2}}$ which is analytic in ${\mathbb{C}^{-}}$ can be defined as
$$
{{P}_{2}}(x,\zeta)=\left( \begin{matrix}
   {[\mu_{-}^{-1}]^{1}}  \\
   {[\mu_{+}^{-1}]^{2}}  \\
   {[\mu_{+}^{-1}]^{3}}  \\
\end{matrix} \right)(x,\zeta).\eqno(14)
$$
Analogous to ${{P}_{1}}$, the very large-$\zeta$ asymptotic behavior of ${{P}_{2}}$ turns out to be
$
{{P}_{2}}\to \mathbb{I}$ as $\zeta \in {\mathbb{C}^{-}}\to \infty .
$

Inserting (5) into (8) gives rise to
\begin{equation*}
({{[{\mu_{-}}]}_{1}},{{[{\mu_{-}}]}_{2}},{{[{\mu_{-}}]}_{3}})=({{[{\mu_{+}}]}_{1}},{{[{\mu_{+}}]}_{2}},{{[{\mu_{+}}]}_{3}})\left( \begin{matrix}
   {{s}_{11}} & {{s}_{12}}{{\text{e}}^{-i\zeta x}} & {{s}_{13}}{{\text{e}}^{-i\zeta x}}  \\
   {{s}_{21}}{{\text{e}}^{i\zeta x}} & {{s}_{22}} & {{s}_{23}}  \\
   {{s}_{31}}{{\text{e}}^{i\zeta x}} & {{s}_{32}} & {{s}_{33}}  \\
\end{matrix} \right),
\end{equation*}
from which we have
\begin{equation*}
{{[{{\mu}_{-}}]}_{1}}={{s}_{11}}{{[{{\mu}_{+}}]}_{1}}+{{s}_{21}}{{\text{e}}^{i\zeta x}}{{[{{\mu}_{+}}]}_{2}}+{{s}_{31}}{{\text{e}}^{i\zeta x}}{{[{{\mu}_{+}}]}_{3}}.
\end{equation*}
Thus, ${{P}_{1}}$ reads
\begin{equation*}
{{P}_{1}}=({{[{\mu_{-}}]}_{1}},{{[{\mu_{+}}]}_{2}},{{[{\mu_{+}}]}_{3}})=({{[{\mu_{+}}]}_{1}},{{[{\mu_{+}}]}_{2}},{{[{\mu_{+}}]}_{3}})\left( \begin{matrix}
   {{s}_{11}} & 0 & 0  \\
   {{s}_{21}}{{\text{e}}^{i\zeta x}} & 1 & 0  \\
   {{s}_{31}}{{\text{e}}^{i\zeta x}} & 0 & 1  \\
\end{matrix} \right).
\end{equation*}

On the other hand, substituting (12) into (13), we derive
\begin{equation*}
\left( \begin{matrix}
   {[\mu_{-}^{-1}]^{1}}  \\
   {[\mu_{-}^{-1}]^{2}}  \\
   {[\mu_{-}^{-1}]^{3}}  \\
\end{matrix} \right)=\left( \begin{matrix}
   {{r}_{11}} & {{r}_{12}}{{\text{e}}^{-i\zeta x}} & {{r}_{13}}{{\text{e}}^{-i\zeta x}}  \\
   {{r}_{21}}{{\text{e}}^{i\zeta x}} & {{r}_{22}} & {{r}_{23}}  \\
   {{r}_{31}}{{\text{e}}^{i\zeta x}} & {{r}_{32}} & {{r}_{33}}  \\
\end{matrix} \right)
\left( \begin{matrix}
   {[\mu_{+}^{-1}]^{1}}  \\
   {[\mu_{+}^{-1}]^{2}}  \\
   {[\mu_{+}^{-1}]^{3}}  \\
\end{matrix} \right),
\end{equation*}
from which ${[\mu_{-}^{-1}]^{1}}$ is of the form
\begin{equation*}
{[\mu_{-}^{-1}]^{1}}={{r}_{11}}{[\mu_{+}^{-1}]^{1}}+{{r}_{12}}{{\text{e}}^{-i\zeta x}}{[\mu_{+}^{-1}]^{2}}+{{r}_{13}}{{\text{e}}^{-i\zeta x}}{[\mu_{+}^{-1}]^{3}}.
\end{equation*}
Subsequently, ${{P}_{2}}$ is represented as
\begin{equation*}
{{P}_{2}}=\left( \begin{matrix}
   {[\mu_{-}^{-1}]^{1}}  \\
   {[\mu_{+}^{-1}]^{2}}  \\
   {[\mu_{+}^{-1}]^{3}}  \\
\end{matrix} \right)=\left( \begin{matrix}
   {{r}_{11}} & {{r}_{12}}{{\text{e}}^{-i\zeta x}} & {{r}_{13}}{{\text{e}}^{-i\zeta x}}  \\
   0 & 1 & 0  \\
   0 & 0 & 1  \\
\end{matrix} \right)\left( \begin{matrix}
   {[\mu_{+}^{-1}]^{1}}  \\
   {[\mu_{+}^{-1}]^{2}}  \\
   {[\mu_{+}^{-1}]^{3}}  \\
\end{matrix} \right).
\end{equation*}

With two matrix functions ${{P}_{1}}$ and ${{P}_{2}}$ which are analytic in ${\mathbb{C}^{+}}$ and ${\mathbb{C}^{-}}$ respectively in hand,
we are ready to describe a Riemann-Hilbert problem for the coupled Hirota system (1).
After denoting that the limit of ${{P}_{1}}$ is ${{P}^{+}}$ as $\zeta\in {\mathbb{C}^{+}}\rightarrow\mathbb{R}$ and
the limit of ${{P}_{2}}$ is ${{P}^{-}}$ as $\zeta \in {\mathbb{C}^{-}}\rightarrow\mathbb{R}$, a Riemann-Hilbert problem desired can be presented below
$$
{{P}^{-}}(x,\zeta){{P}^{+}}(x,\zeta)=\left( \begin{matrix}
   1 & {{r}_{12}}{{\text{e}}^{-i\zeta x}} & {{r}_{13}}{{\text{e}}^{-i\zeta x}}  \\
   {{s}_{21}}{{\text{e}}^{i\zeta x}} & 1 & 0  \\
   {{s}_{31}}{{\text{e}}^{i\zeta x}} & 0 & 1  \\
\end{matrix} \right)\eqno(15)
$$
with its canonical normalization conditions given by
\begin{eqnarray*}
  &{{P}_{1}}(x,\zeta)\to \mathbb{I},\quad \zeta \in {\mathbb{C}^{+}}\to \infty , \\
  &{{P}_{2}}(x,\zeta)\to \mathbb{I},\quad \zeta \in {\mathbb{C}^{-}}\to \infty ,
\end{eqnarray*}
and ${{r}_{11}}{{s}_{11}}+{{r}_{12}}{{s}_{21}}+{{r}_{13}}{{s}_{31}}=1$.

\section{Soliton solutions}
In this section, we will construct soliton solutions to system (1) based on the Riemann-Hilbert problem presented above.
We now posit the Riemann-Hilbert problem (15) to be irregular, which signifies that both $\det {{P}_{1}}$ and $\det {{P}_{2}}$ are in possession of some zeros in analytic domains of their own.
According to the definitions of ${{P}_{1}}$ and ${{P}_{2}}$, we have
\begin{align}
& \det {{P}_{1}}(\zeta)={{s}_{11}}(\zeta),\quad \zeta \in {\mathbb{C}^{+}},  \nonumber \\
& \det {{P}_{2}}(\zeta)={{r}_{11}}(\zeta),\quad \zeta \in {\mathbb{C}^{-}},  \nonumber
\end{align}
which enable us to see that
$\det {{P}_{1}}$ and $\det {{P}_{2}}$ possess the same zeros as ${s}_{11}$ and ${r}_{11}$ respectively.

With above analysis, we now study the characteristic feature of zeros.
We note that the potential matrix $Q$ has the symmetry relation
$
Q^{\dagger }=-Q.
$
Here $\dagger$ stands for the Hermitian of a matrix. On basis of this relation, we deduce
$$
\mu_{\pm }^{\dagger }({{\zeta}^{*}})=\mu_{\pm }^{-1}(\zeta).\eqno(16)
$$
In order to ease discussion, we introduce two matrices ${{H}_{1}}=\text{diag}(1,0,0)$ and ${{H}_{2}}=\text{diag}(0,1,1)$, which permits us to express (9) and (14) as
\begin{subequations}
\begin{align}
&{{P}_{1}}={\mu_{-}}{{H}_{1}}+{\mu_{+}}{{H}_{2}},\tag{17a}\\
&{{P}_{2}}={{H}_{1}}\mu_{-}^{-1}+{{H}_{2}}\mu_{+}^{-1}.\tag{17b}
\end{align}
\end{subequations}
By taking the Hermitian of expression (17a) and using the relation (16), we find
$$
P_{1}^{\dagger }({{\zeta}^{*}})={{P}_{2}}(\zeta),\quad \zeta \in {\mathbb{C}^{-}},\eqno(18)
$$
and the involution property of the scattering matrix
$
{{S}^{\dagger }}({{\zeta}^{*}})={{S}^{-1}}(\zeta),
$
which follows at once
$$
s_{11}^{*}({{\zeta}^{*}})={{r}_{11}}(\zeta),\quad \zeta \in {\mathbb{C}^{-}}.\eqno(19)
$$
This equality implies that each zero $\pm {{\zeta}_{k}}$ of ${{s}_{11}}$ generates each zero $\pm \zeta_{k}^{*}$ of ${{r}_{11}}$ correspondingly.
Therefore, our assumption is that $\det {{P}_{1}}$ has $N$ simple zeros $\{{{\zeta}_{j}}\}_{j=1}^{N}$ in ${\mathbb{C}^{+}}$
and $\det {{P}_{2}}$ has $N$ simple zeros $\{{{\hat{\zeta}}_{j}}\}_{j=1}^{N}$ in ${\mathbb{C}^{-}}$, where
${{\hat{\zeta}}_{l}}=\zeta_{l}^{*},$ $ 1\le l\le N.$
The full set of the discrete scattering data is composed of these zeros and the nonzero column vectors ${{\upsilon}_{j}}$ as well as row vectors ${{\hat{\upsilon}}_{j}}$, which satisfy the following equations
\begin{subequations}
\begin{align}
&{{P}_{1}}({{\zeta}_{j}}){{\upsilon}_{j}}=0,\tag{20a}\\
&{{\hat{\upsilon}}_{j}}{{P}_{2}}({{\hat{\zeta}}_{j}})=0.\tag{20b}
\end{align}
\end{subequations}

Taking the Hermitian of (20a) and using (18) as well as comparing with (20b), we find that the eigenvectors fulfill the relation
$$
{{\hat{\upsilon}}_{j}}=\upsilon_{j}^{\dagger },\quad 1\le j\le N.\eqno(21)
$$
Differentiating (20a) with respect to $x$ and $t$ respectively and taking advantage of (4), we get
\begin{equation*}
\begin{aligned}
 & {{P}_{1}}({{\zeta}_{j}})\left( \frac{\partial {{\upsilon }_{j}}}{\partial x}-\frac{i}{2}{{\zeta}_{j}}\sigma {{\upsilon}_{j}} \right)=0, \\
 & {{P}_{1}}({{\zeta}_{j}})\left( \frac{\partial {{\upsilon }_{j}}}{\partial t}+\left( \frac{i}{2}\zeta _{j}^{3}\varepsilon +\zeta_{j}^{2}{{A}_{2}} \right)\sigma {{\upsilon}_{j}} \right)=0,
\end{aligned}
\end{equation*}
from which we find
\begin{equation*}
{{\upsilon }_{j}}={{\text{e}}^{\frac{i}{2}{{\zeta}_{j}}\sigma x-\big( \frac{i}{2}\zeta_{j}^{3}\varepsilon +\zeta_{j}^{2}{{A}_{2}} \big)\sigma t}}{{\upsilon }_{j,0}},\quad 1\le j\le N.
\end{equation*}
Here ${{\upsilon}_{j,0}}$ are independent of the variables $x$ and $t$. In consideration of the relation (21), we thus have
\begin{equation*}
{{\hat{\upsilon }}_{j}}=\upsilon _{j,0}^{\dagger }{{\text{e}}^{-\frac{i}{2}\zeta_{j}^{*}\sigma x-\big( -\frac{i}{2}\zeta_{j}^{*3}\varepsilon +\zeta _{j}^{*2}{{A}_{2}} \big)\sigma t}},\quad 1\le j\le N.
\end{equation*}

It is noted that the Riemann-Hilbert problem (15) corresponds to the reflectionless case, namely, ${{s}_{21}}={{s}_{31}}=0$. The expression of solution
to the problem (15) reads
\begin{subequations}
\begin{align}
& {{P}_{1}}(\zeta)=\mathbb{I}-\sum\limits_{k=1}^{N}{\sum\limits_{j=1}^{N}{\frac{{{\upsilon}_{k}}{{{\hat{\upsilon}}}_{j}}{{\big({{M}^{-1}}\big)}_{kj}}}{\zeta -{{{\hat{\zeta}}}_{j}}}}},\tag{22a}\\
& {{P}_{2}}(\zeta)=\mathbb{I}+\sum\limits_{k=1}^{N}{\sum\limits_{j=1}^{N}{\frac{{{\upsilon}_{k}}{{{\hat{\upsilon}}}_{j}}{{\big({{M}^{-1}}\big)}_{kj}}}{\zeta-{{\zeta }_{k}}}}},\tag{22b}
\end{align}
\end{subequations}
in which $M$ is a $N\times N$ matrix defined as
\begin{equation*}
M=({{M}_{kj}})_{N\times N}=\left(\frac{{\hat{\upsilon}_{k}}{{{{\upsilon}}}_{j}}}{{{\zeta}_{j}}-{{{\hat{\zeta}}}_{k}}}\right)_{N\times N},\quad 1\le k,j\le N,
\end{equation*}
and ${{\big({{M}^{-1}}\big)}_{kj}}$ stands for the $(k,j)$-entry of the inverse matrix of $M$. From expression (22a), we can immediately obtain
\begin{equation*}
P_{1}^{(1)}=-\sum\limits_{k=1}^{N}{\sum\limits_{j=1}^{N}{{{\upsilon}_{k}}{{{\hat{\upsilon}}}_{j}}{{\big({{M}^{-1}}\big)}_{kj}}}}.
\end{equation*}

In what follows, we shall retrieve the potential functions $q_{1}$ and $q_{2}$. Expanding ${{P}_{1}}(\zeta)$ at large-$\zeta$ as
\begin{equation*}
{{P}_{1}}(\zeta)=\mathbb{I}+\frac{P_{1}^{(1)}}{\zeta}+\frac{P_{1}^{(2)}}{{{\zeta}^{2}}}+O\bigg( \frac{1}{{{\zeta}^{3}}} \bigg),\quad \zeta \to \infty .
\end{equation*}
and putting it into (4a) gives
\begin{equation*}
Q=\frac{i}{2{{k}_{1}}}\big[\sigma ,P_{1}^{(1)}\big].
\end{equation*}
Hence, the potential functions can be reconstructed as
\begin{equation*}
\begin{aligned}
 & {{q}_{1}}=-\frac{i}{{{k}_{1}}}{{\big(P_{1}^{(1)}\big)}_{12}}, \\
 & {{q}_{2}}=-\frac{i}{{{k}_{1}}}{{\big(P_{1}^{(1)}\big)}_{13}}, \\
\end{aligned}
\end{equation*}
where ${{\big(P_{1}^{(1)}\big)}_{12}}$ and ${{\big(P_{1}^{(1)}\big)}_{13}}$ are respectively the (1,2)- and (1,3)-entries of matrix $P_{1}^{(1)}$.

As a result, setting the nonzero vectors ${{\upsilon}_{k,0}}={({{\alpha }_{k}},{{\beta }_{k}},{{\gamma}_{k}})^\textrm{T}}$ and ${{\theta }_{k}}=\frac{i}{2}{{\zeta}_{k}}x-\big( \frac{i}{2}\zeta_{k}^{3}\varepsilon +\zeta_{k}^{2}{{A}_{2}}\big)t$,\ $\operatorname{Im}({{\zeta}_{k}})>0,1\le k\le N$,
the expression of general $N$-soliton solution for system (1) is of the form
\begin{subequations}
\begin{align}
 & {{q}_{1}}=\frac{i}{{{k}_{1}}}{{\sum\limits_{k=1}^{N}{\sum\limits_{j=1}^{N}{{{\alpha }_{k}}\beta _{j}^{*}{{\text{e}}^{{-{\theta }_{k}}+\theta _{j}^{*}}}{{\big({{M}^{-1}}\big)}_{kj}}}}}},\tag{23a} \\
 & {{q}_{2}}=\frac{i}{{{k}_{1}}}{{\sum\limits_{k=1}^{N}{\sum\limits_{j=1}^{N}{{{\alpha }_{k}}\gamma _{j}^{*}{{\text{e}}^{{-{\theta }_{k}}+\theta _{j}^{*}}}{{\big({{M}^{-1}}\big)}_{kj}}}}}},\tag{23b}
\end{align}
\end{subequations}
where
\[
{{M}_{kj}}=\frac{\alpha _{k}^{*}{{\alpha }_{j}}{{\text{e}}^{-\theta _{k}^{*}-{{\theta }_{j}}}}+\beta _{k}^{*}{{\beta }_{j}}{{\text{e}}^{\theta _{k}^{*}+{{\theta }_{j}}}}+\gamma _{k}^{*}{{\gamma }_{j}}{{\text{e}}^{\theta _{k}^{*}+{{\theta }_{j}}}}}{{{\zeta}_{j}}-\zeta_{k}^{*}},\quad 1\le k,j\le N.
\]

At the end of this section, the one-bright-soliton solution will be our main concern. For the case of $N=1$, the corresponding one-bright-soliton solution can be readily given as
\begin{subequations}\begin{align}
 & {{q}_{1}}=\frac{i}{{{k}_{1}}}{{\alpha }_{1}}\beta _{1}^{*}{{\text{e}}^{-{{\theta }_{1}}+\theta _{1}^{*}}}\frac{{{\zeta}_{1}}-\zeta_{1}^{*}}{{{\left| {{\alpha }_{1}} \right|}^{2}}{{\text{e}}^{-{{\theta }_{1}}-\theta _{1}^{*}}}+\big({{\left| {{\beta }_{1}} \right|}^{2}}+{{\left| {{\gamma }_{1}} \right|}^{2}}\text{\big)}{{\text{e}}^{{{\theta }_{1}}+\theta _{1}^{*}}}},\tag{24a} \\
 & {{q}_{2}}=\frac{i}{{{k}_{1}}}{{\alpha }_{1}}\gamma _{1}^{*}{{\text{e}}^{-{{\theta }_{1}}+\theta _{1}^{*}}}\frac{{{\zeta }_{1}}-\zeta_{1}^{*}}{{{\left| {{\alpha }_{1}} \right|}^{2}}{{\text{e}}^{-{{\theta }_{1}}-\theta _{1}^{*}}}+\big({{\left| {{\beta }_{1}} \right|}^{2}}+{{\left| {{\gamma }_{1}} \right|}^{2}}\text{\big)}{{\text{e}}^{{{\theta }_{1}}+\theta _{1}^{*}}}}.\tag{24b}
\end{align}\end{subequations}
Here
${{\theta }_{1}}=\frac{i}{2}{{\zeta}_{1}}x-\big( \frac{i}{2}\zeta_{1}^{3}\varepsilon +\zeta_{1}^{2}{{A}_{2}}\big)t$.
And then, we choose ${{\alpha }_{1}}=1$ and assume ${{\left| {{\beta }_{1}} \right|}^{2}}+{{\left| {{\gamma }_{1}} \right|}^{2}}={{\text{e}}^{2{{\xi }_{1}}}}$ as well as ${{\zeta}_{1}}={{a}_{1}}+i{{b}_{1}}$, which allows the solution (24) to be expressed as
\begin{subequations}\begin{align}
 & {{q}_{1}}=-\frac{\beta _{1}^{*}{{b}_{1}}}{{{k}_{1}}}{{\text{e}}^{-{{\xi }_{1}}}}{{\text{e}}^{-{{\theta }_{1}}+\theta _{1}^{*}}}\text{sech}({{\theta }_{1}}+\theta _{1}^{*}+{{\xi }_{1}}),\tag{25a} \\
 & {{q}_{2}}=-\frac{\gamma _{1}^{*}{{b}_{1}}}{{{k}_{1}}}{{\text{e}}^{-{{\xi }_{1}}}}{{\text{e}}^{-{{\theta }_{1}}+\theta _{1}^{*}}}\text{sech}({{\theta }_{1}}+\theta _{1}^{*}+{{\xi }_{1}}),\tag{25b}
\end{align}\end{subequations}
in which ${{\theta }_{1}}=\frac{i}{2}({{a}_{1}}+i{{b}_{1}})x-\big(\frac{i}{2}{{({{a}_{1}}+i{{b}_{1}})}^{3}}\varepsilon +{{({{a}_{1}}+i{{b}_{1}})}^{2}}{{A}_{2}}\big)t.$

By selecting the involved parameters as ${{k}_{1}}=1,{{\zeta}_{1}}=0.3+0.2i,{{A}_{2}}=1,\varepsilon =1,{{\beta }_{1}}=1,{{\gamma }_{1}}=2,{{\xi }_{1}}=\frac{1}{2}\ln 5,$ we display the localized structures and dynamic behaviors of one-bright-soliton solution (25) in Figures 1, 2 and 3 respectively.

\begin{figure}
\begin{center}
\subfigure[]{\resizebox{0.35\hsize}{!}{\includegraphics*{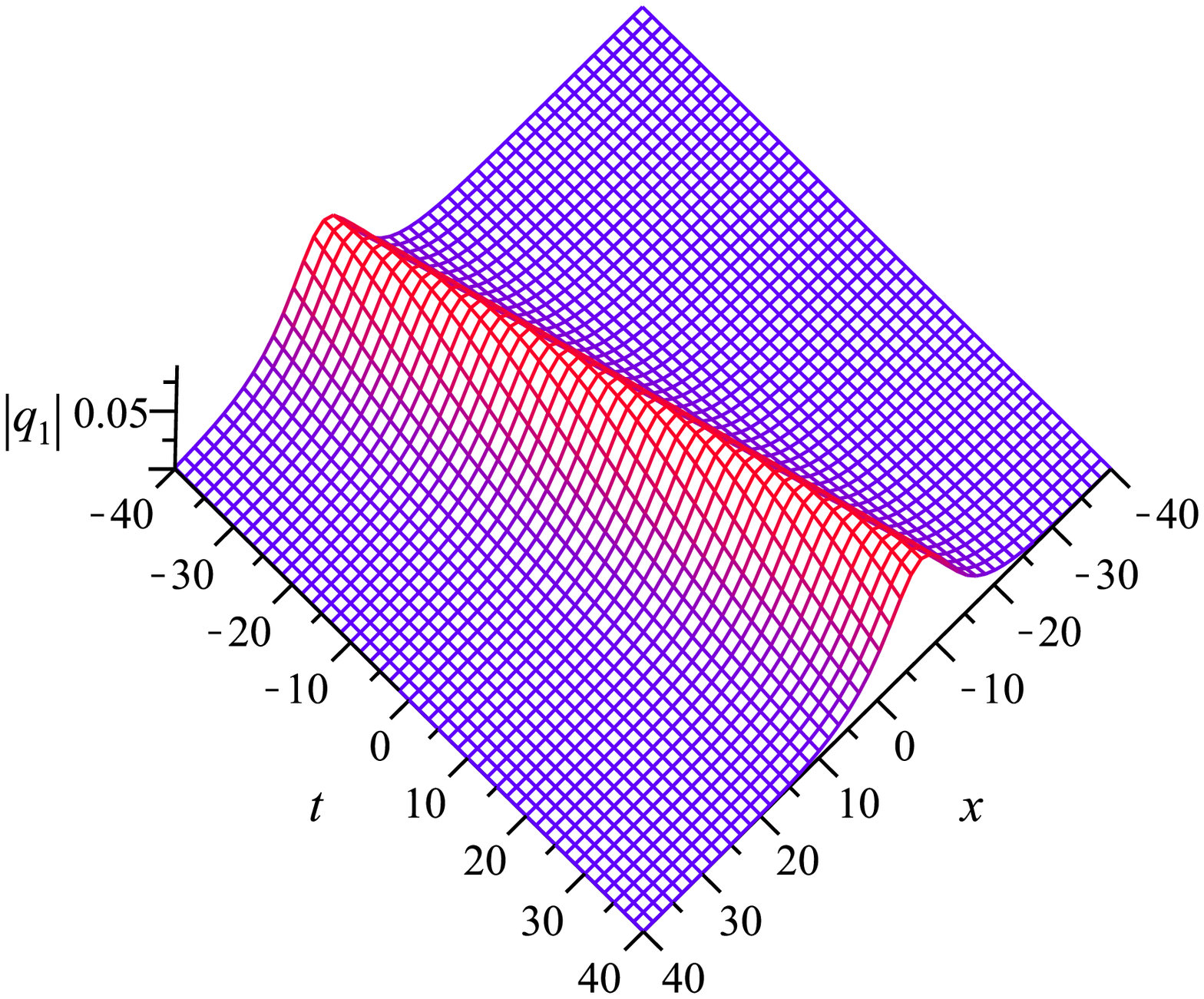}}}
\subfigure[]{\resizebox{0.35\hsize}{!}{\includegraphics*{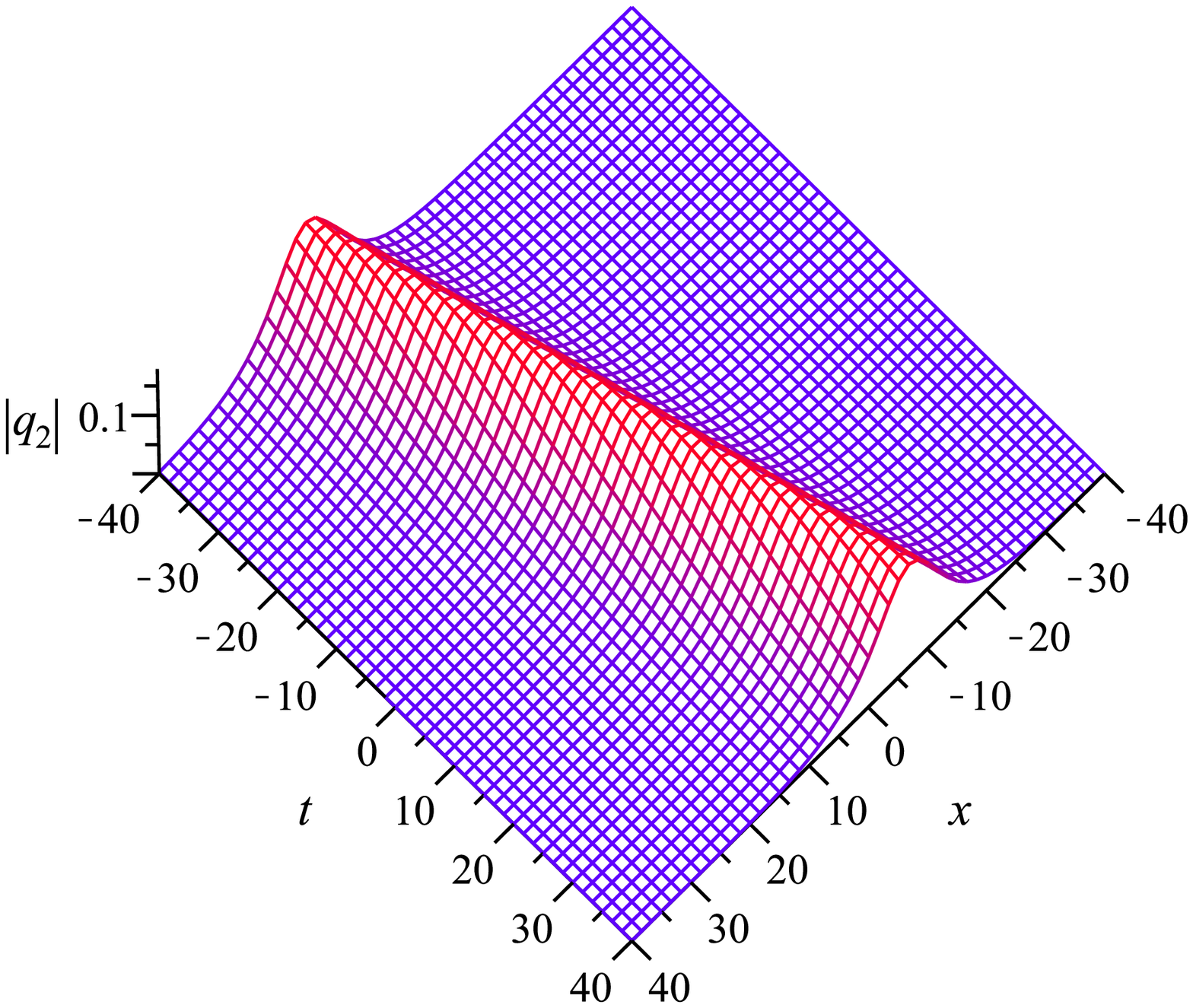}}}
\subfigure[]{\resizebox{0.31\hsize}{!}{\includegraphics*{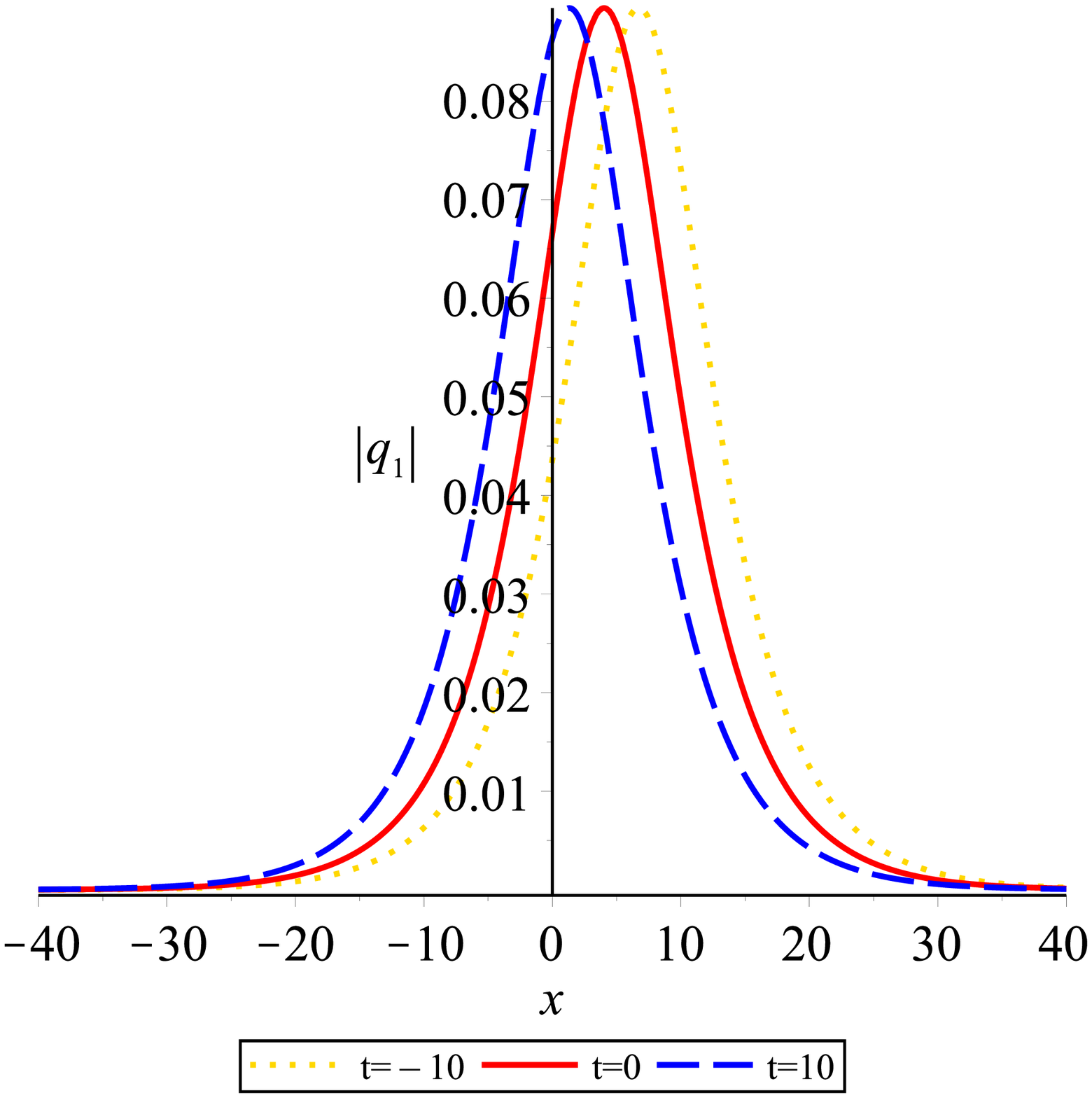}}}
\subfigure[]{\resizebox{0.31\hsize}{!}{\includegraphics*{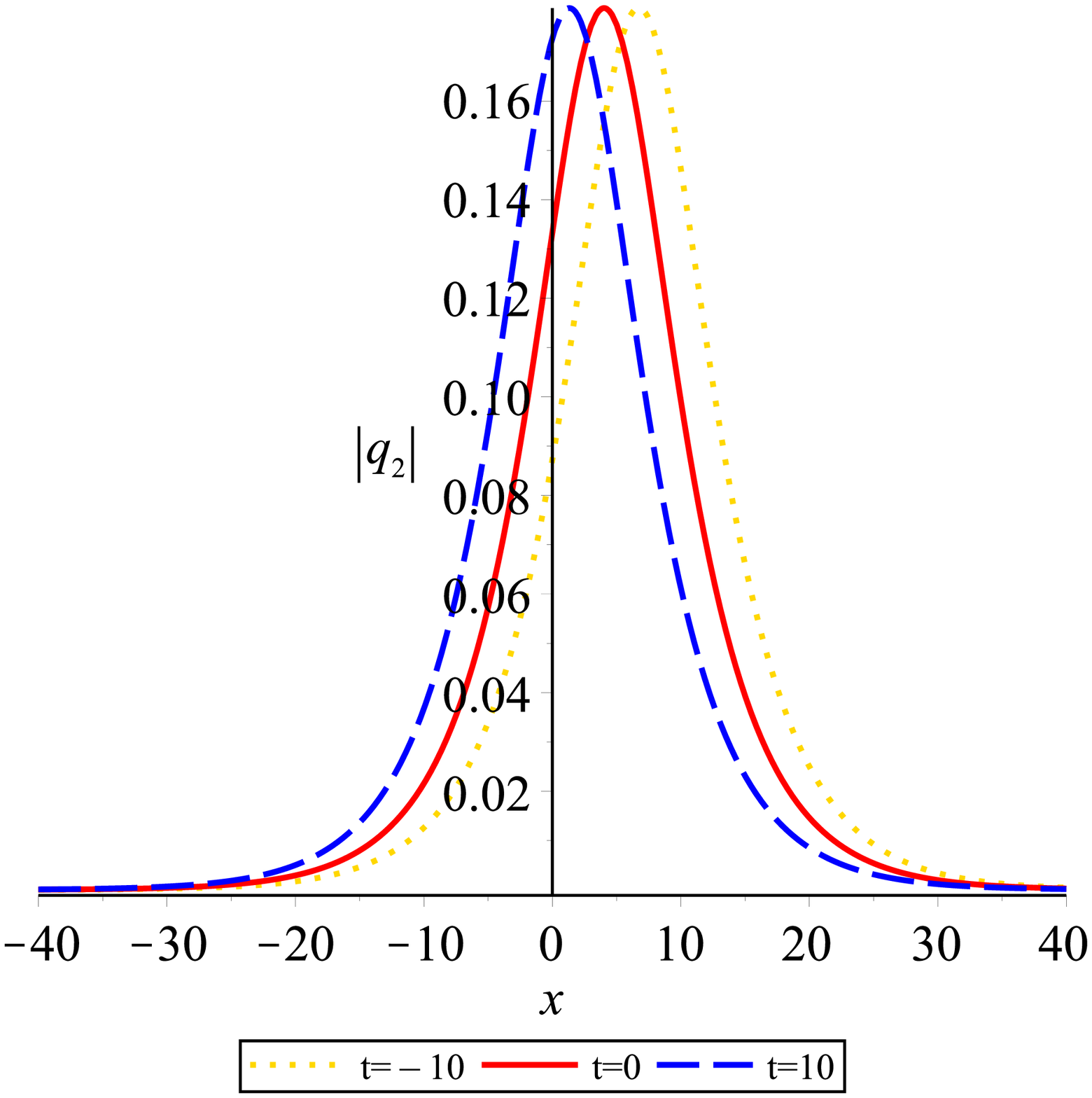}}}
\parbox[c]{12.0cm}{\footnotesize{\bf Figure~1.~~}
One-soliton solution (25). (a) Perspective view of modulus of $q_{1}$; (b) Perspective view of modulus of $q_{2}$; (c) The soliton along the $x$-axis with different time in Fig. 1(a); (d) The soliton along the $x$-axis with different time in Fig. 1(b).}
\end{center}
\addtocounter{subfigure}{-4}
\end{figure}

\begin{figure}
\begin{center}
\subfigure[]{\resizebox{0.35\hsize}{!}{\includegraphics*{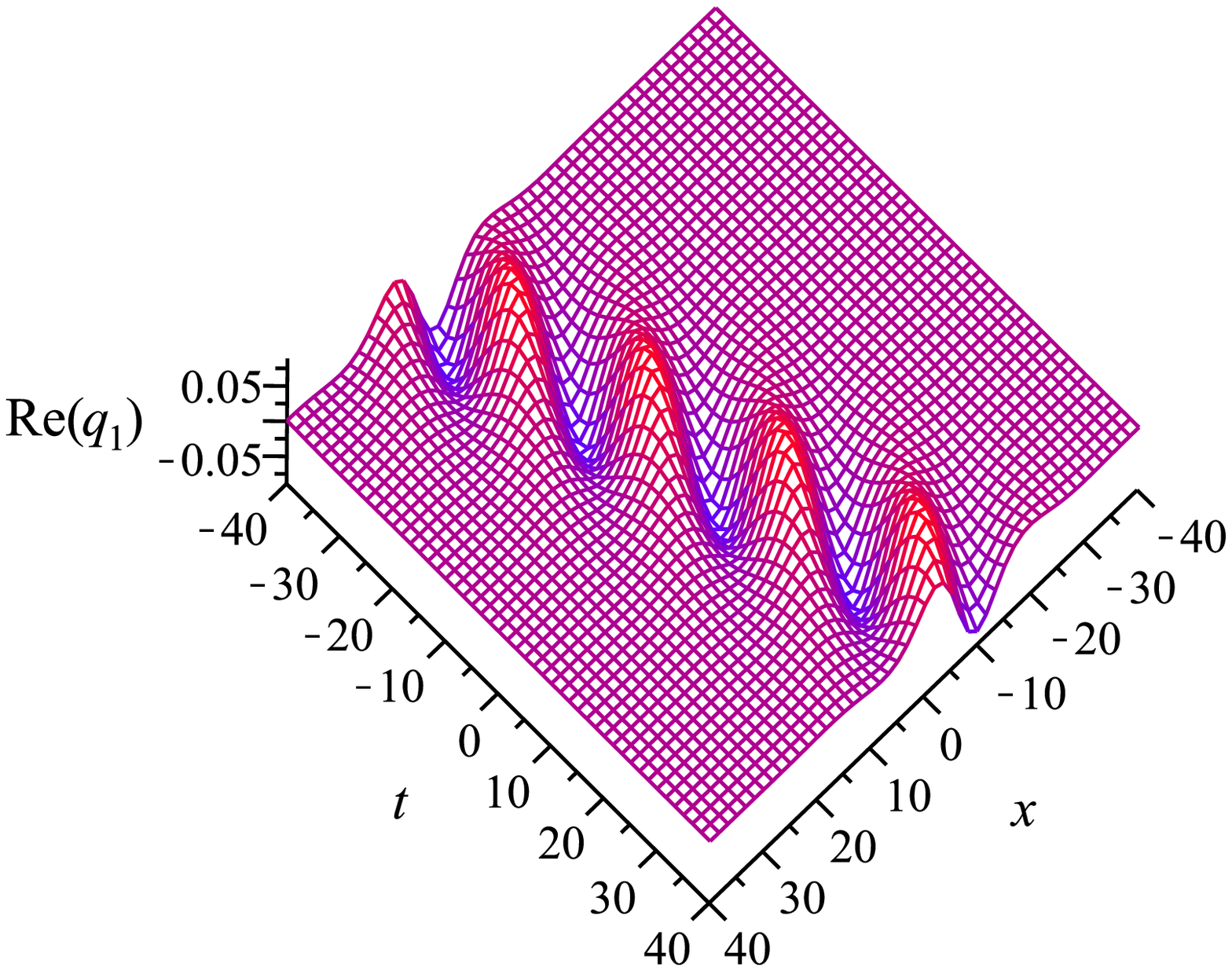}}}
\subfigure[]{\resizebox{0.35\hsize}{!}{\includegraphics*{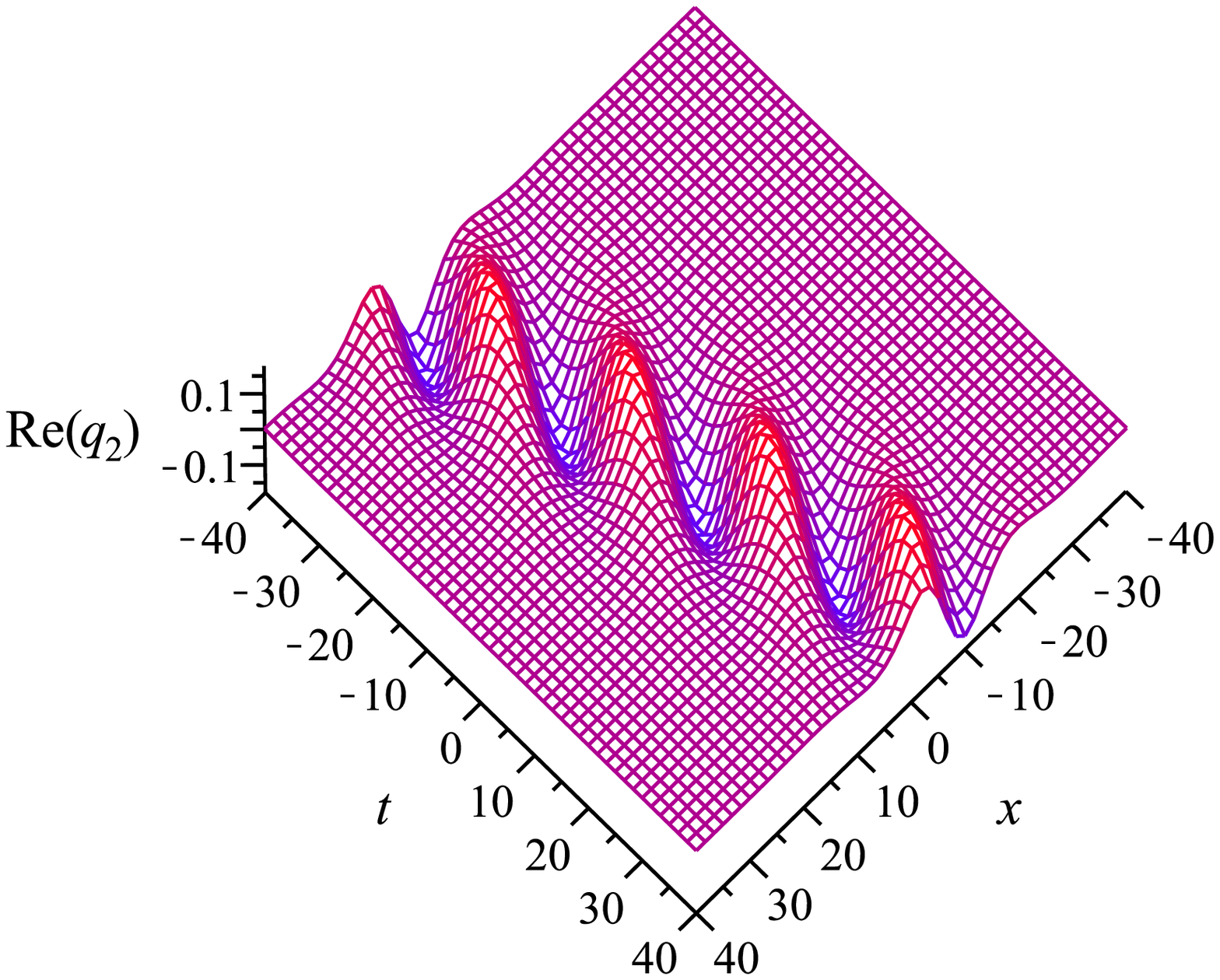}}}
\subfigure[]{\resizebox{0.31\hsize}{!}{\includegraphics*{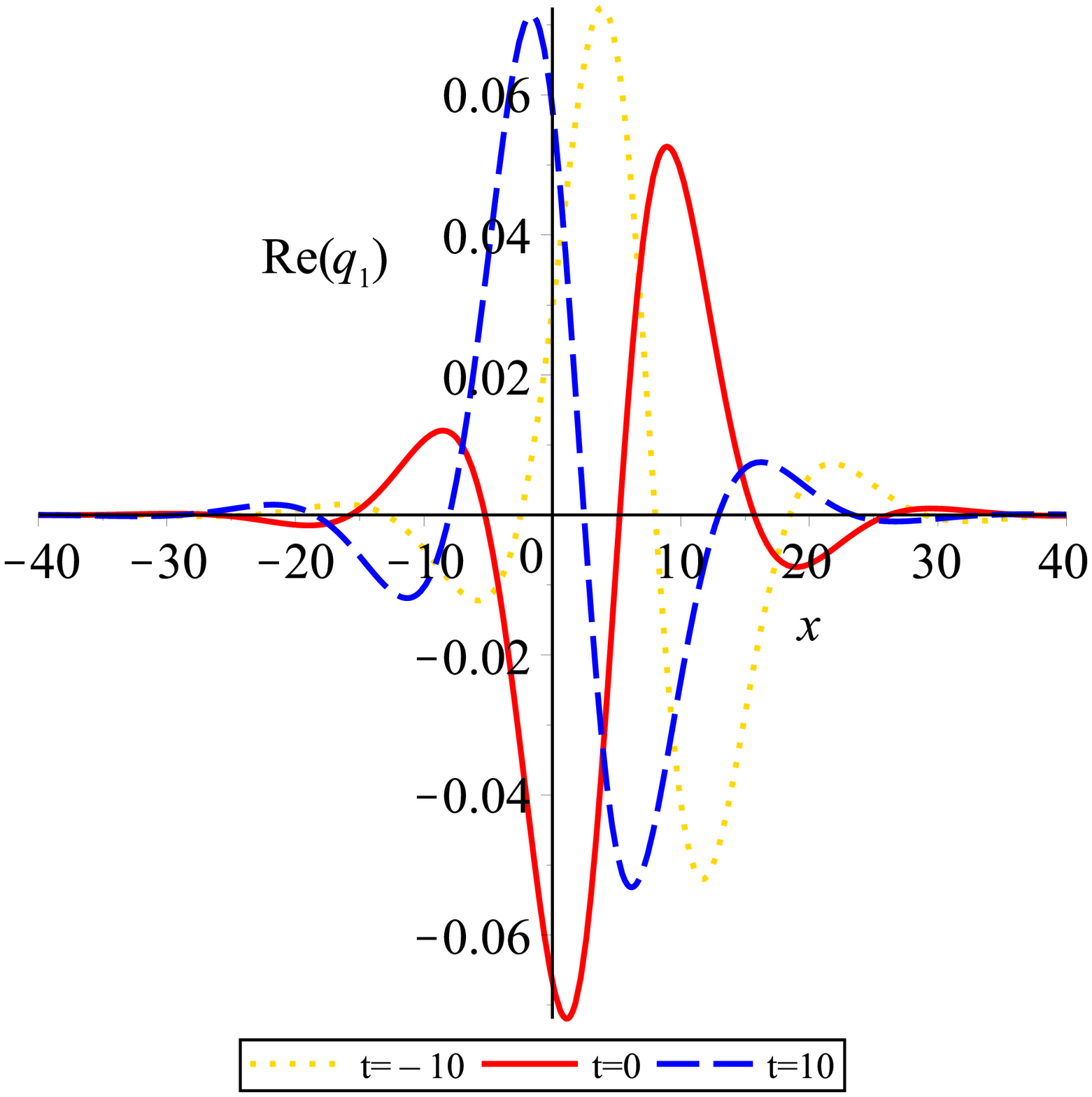}}}
\subfigure[]{\resizebox{0.31\hsize}{!}{\includegraphics*{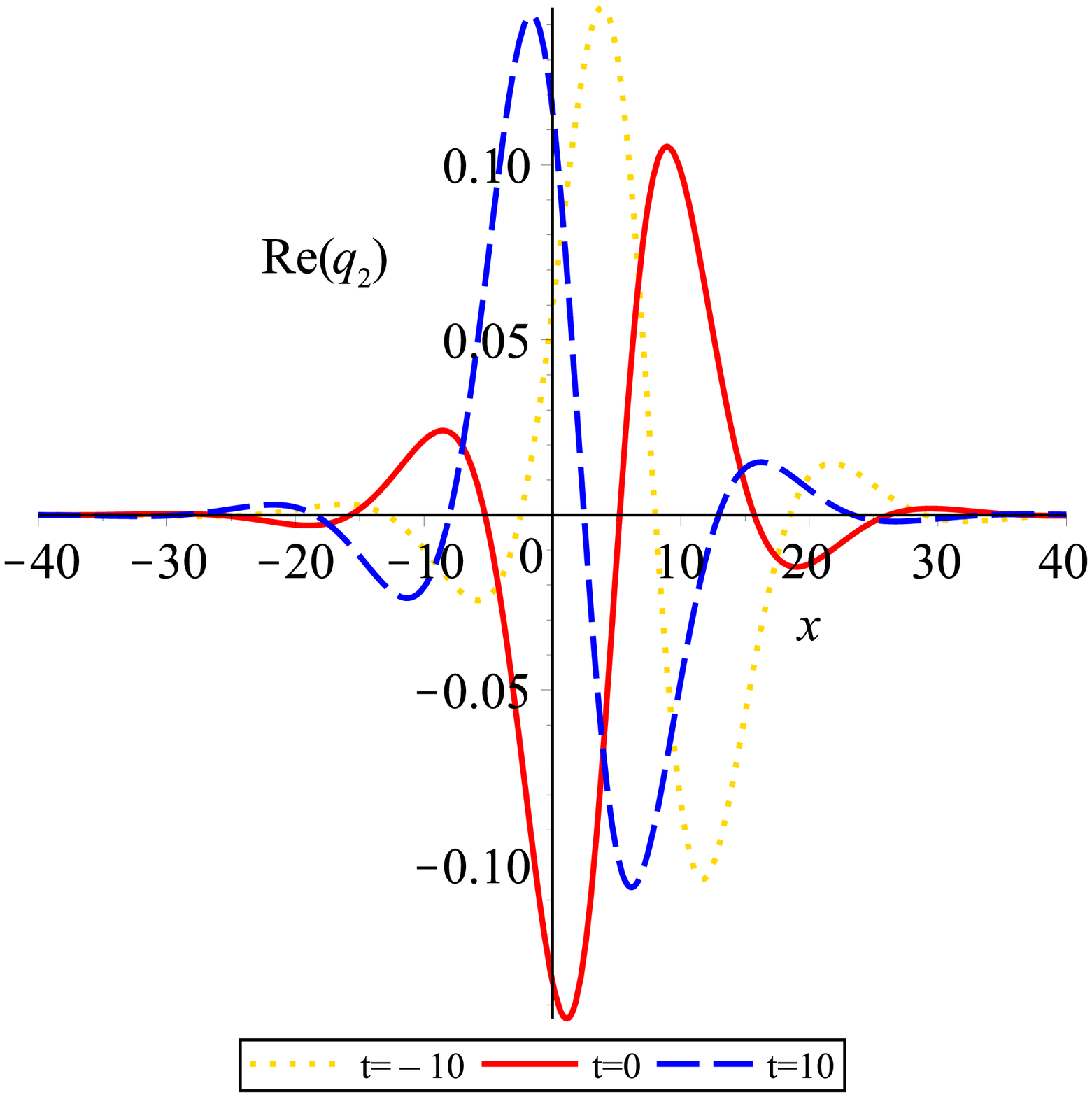}}}
\parbox[c]{12.0cm}{\footnotesize{\bf Figure~2.~~}
One-soliton solution (25). (a) Perspective view of real part of $q_{1}$; (b) Perspective view of real part of $q_{2}$; (c) The soliton along the $x$-axis with different time in Fig. 2(a); (d) The soliton along the $x$-axis with different time in Fig. 2(b).}
\end{center}
\addtocounter{subfigure}{-4}
\end{figure}

\begin{figure}
\begin{center}
\subfigure[]{\resizebox{0.35\hsize}{!}{\includegraphics*{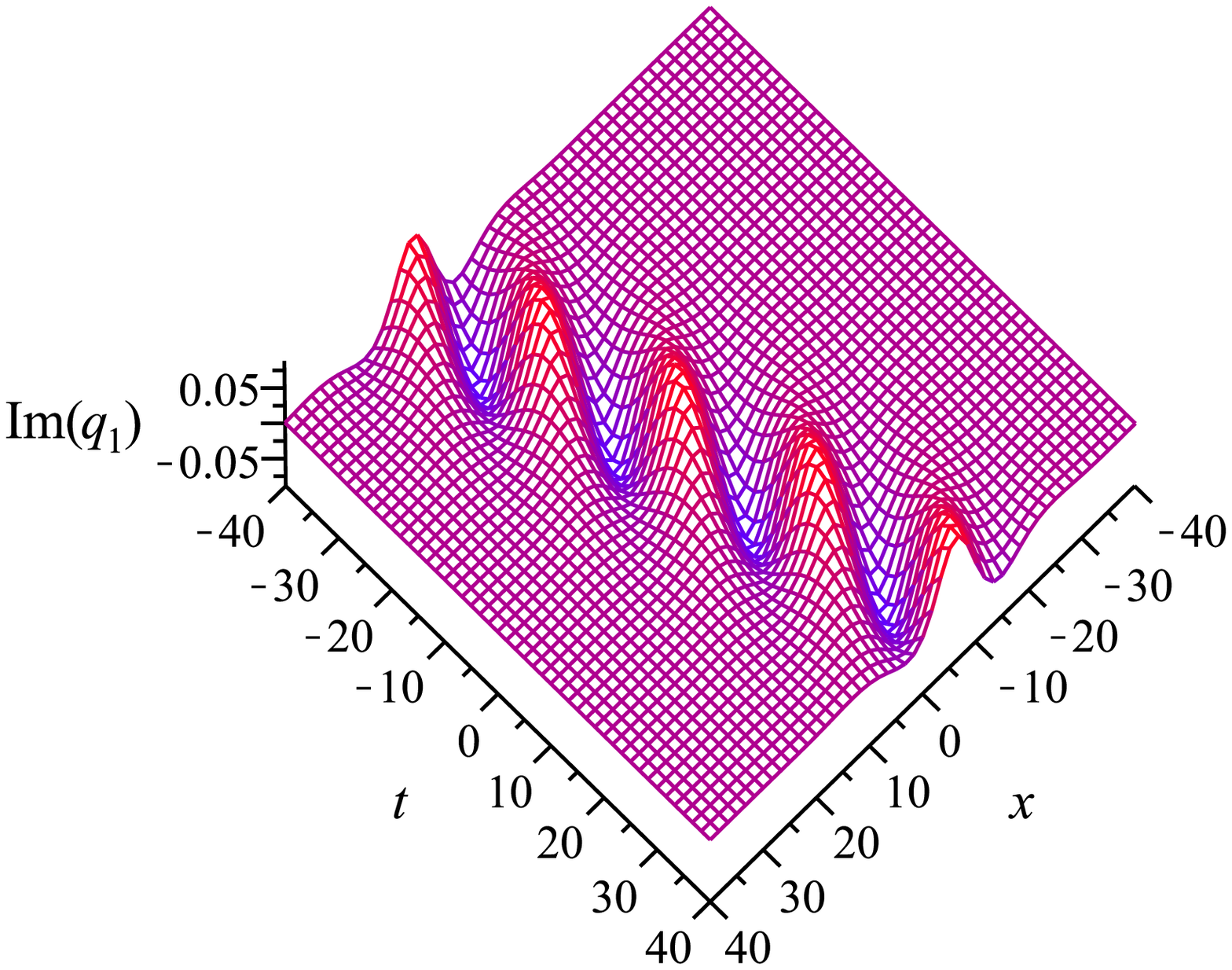}}}
\subfigure[]{\resizebox{0.35\hsize}{!}{\includegraphics*{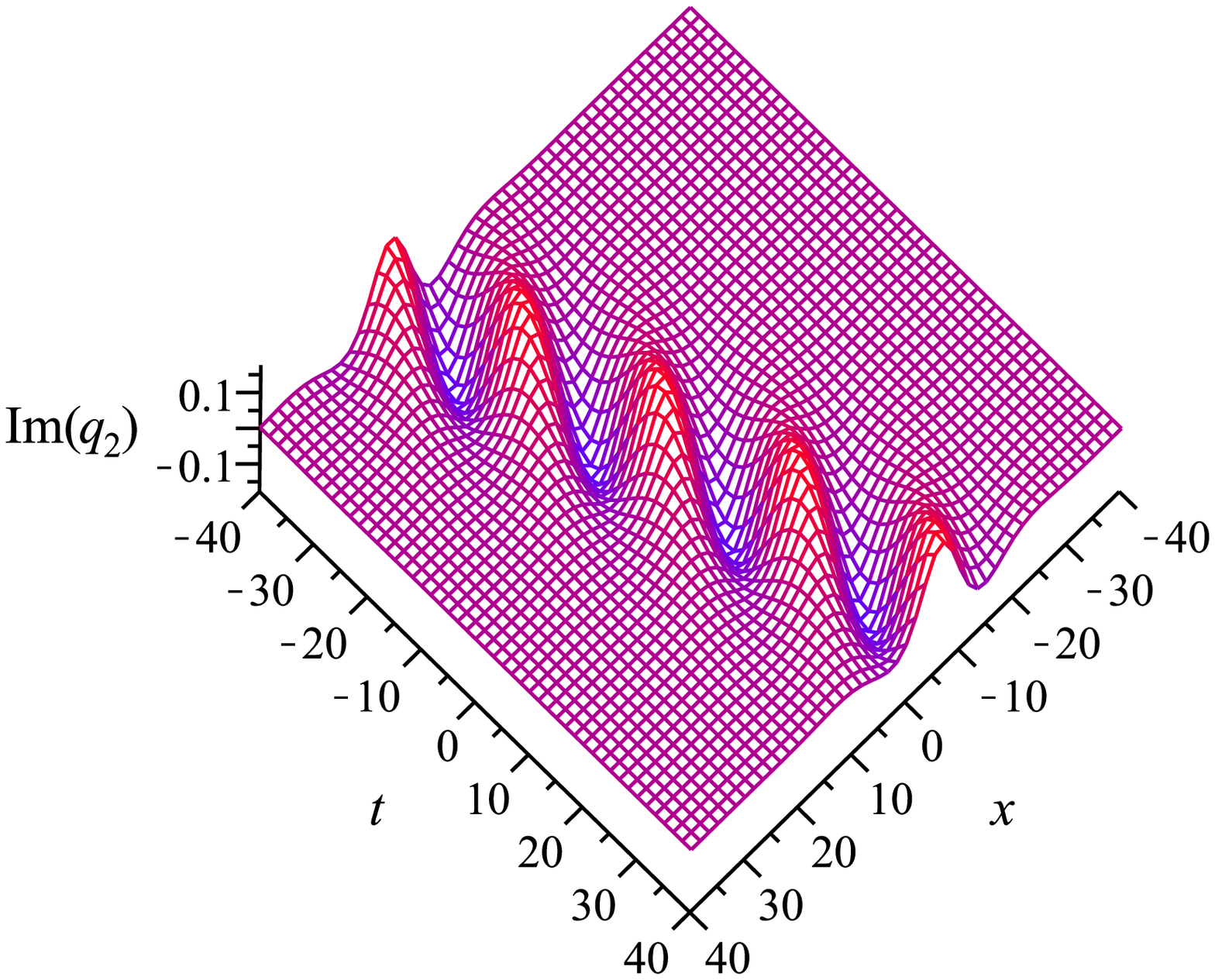}}}
\subfigure[]{\resizebox{0.31\hsize}{!}{\includegraphics*{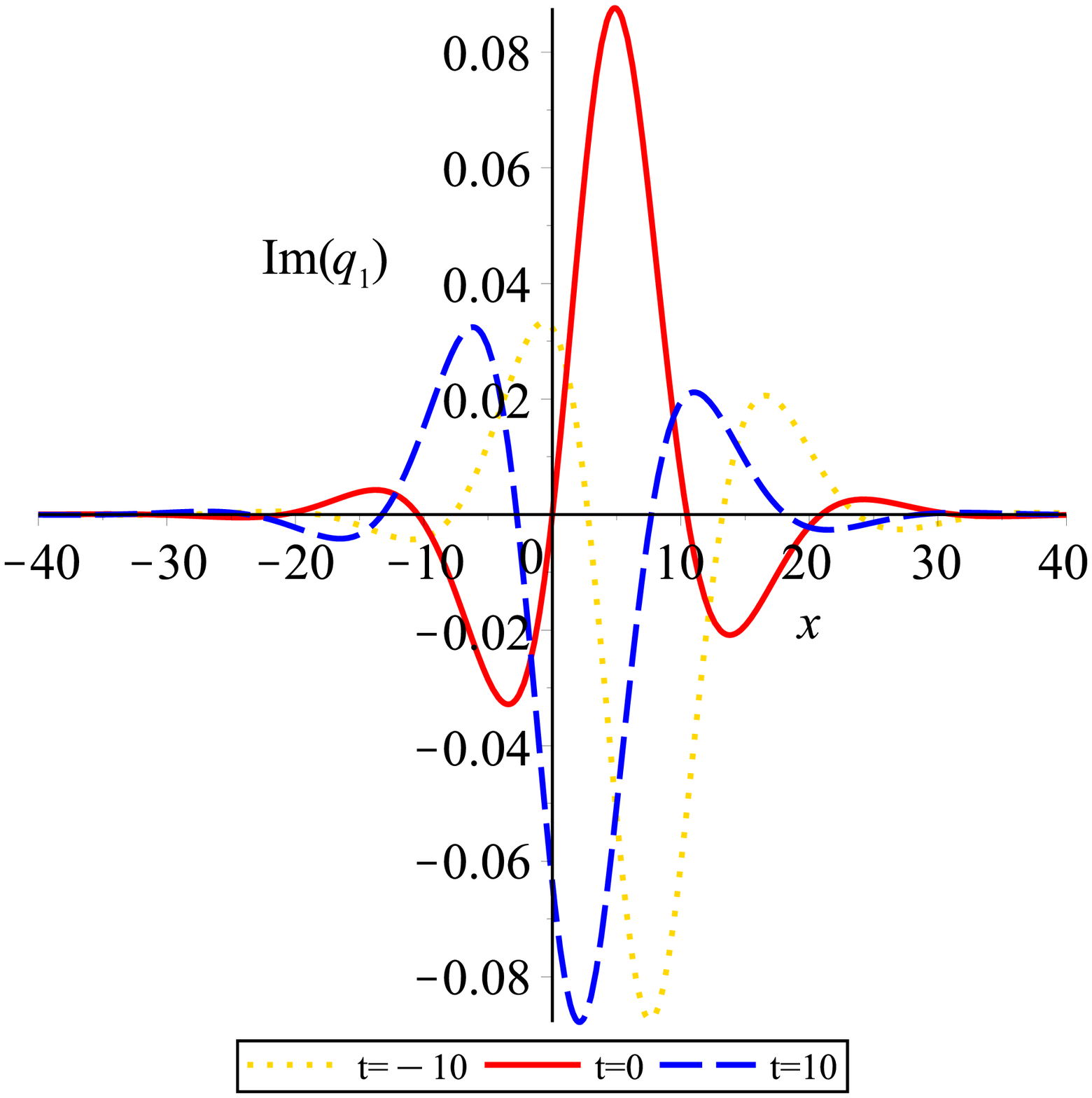}}}
\subfigure[]{\resizebox{0.31\hsize}{!}{\includegraphics*{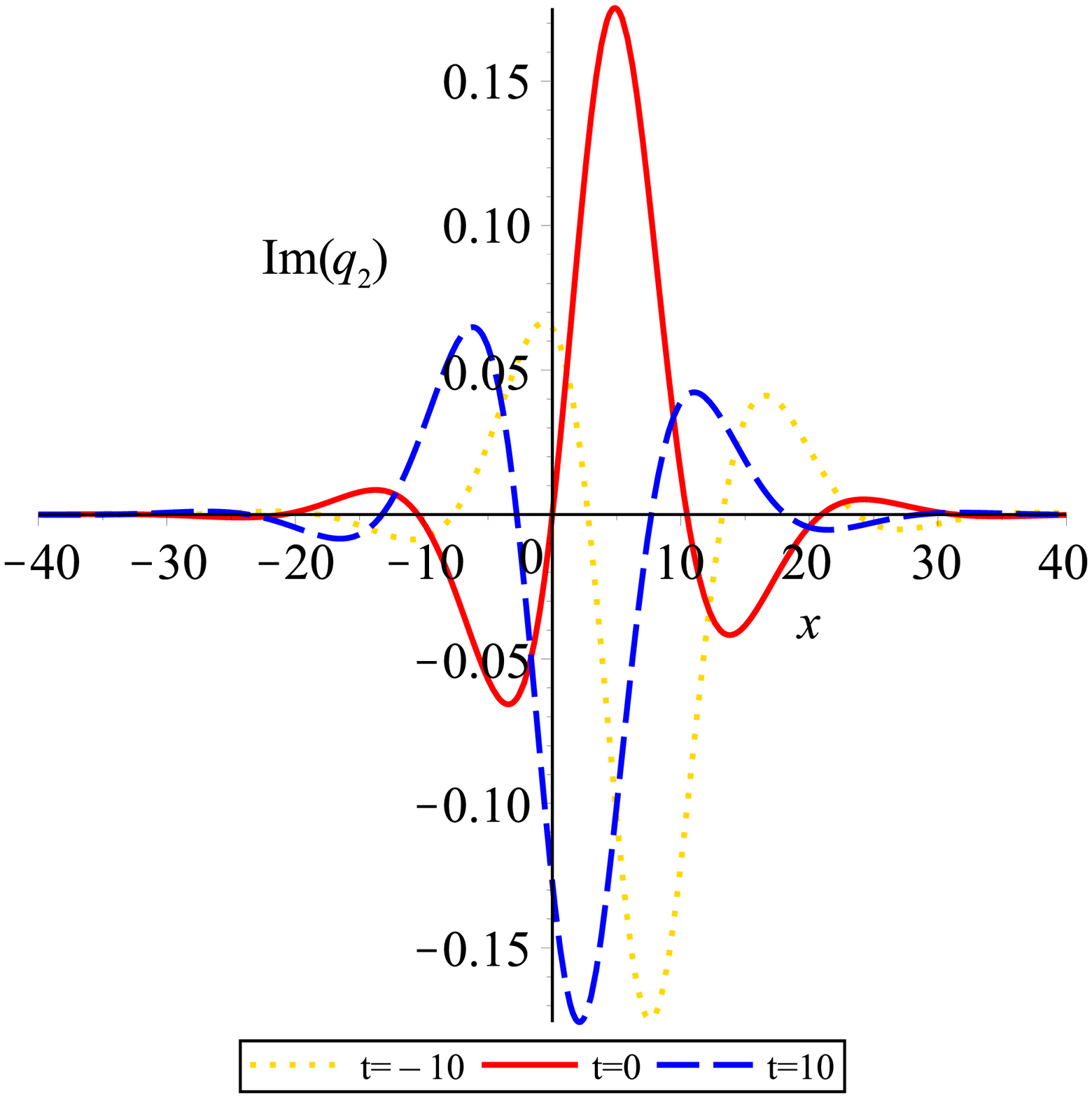}}}
\parbox[c]{12.0cm}{\footnotesize{\bf Figure~3.~~}
One-soliton solution (25). (a) Perspective view of imaginary of $q_{1}$; (b) Perspective view of imaginary of $q_{2}$; (c) The soliton along the $x$-axis with different time in Fig. 3(a); (d) The soliton along the $x$-axis with different time in Fig. 3(b).}
\end{center}
\addtocounter{subfigure}{-4}
\end{figure}
\section{Conclusion}
The aim of the present research was to seek soliton solutions of the coupled Hirota system arising in nonlinear fiber. To this end, the Riemann-Hilbert method was adopted which relays on a Riemann-Hilbert problem. Therefore, the spectral problem of the Lax pair was analyzed and a Riemann-Hilbert problem was worked out. After that, in the framework of the Riemann-Hilbert problem under the reflectionless case, the general $N$-soliton solution for the coupled Hirota system was presented explicitly. Furthermore, we wrote out one-bright-soliton solution and showed its localized structures and dynamic characteristics via some three-dimensional plots and two-dimensional curves with selections of the involved parameters.

\end{document}